\pdfoutput=1
\documentclass[a4paper,12pt]{article}
\usepackage{a4wide}
\usepackage[a4paper, top = 2cm, bottom = 2cm, left = 1cm, right = 1cm]{geometry}
\usepackage[utf8]{inputenc}
\usepackage[T1]{fontenc}
\usepackage{cite}
\usepackage{amssymb}
\usepackage{amsfonts}
\usepackage{graphicx}
\usepackage[section]{placeins}

\newcommand{\beq}{\begin{equation}}
\newcommand{\eeq}{\end{equation}}

\begin{document}
%tytuł i autor
\title{
%\vspace*{-12ex}
Applications of H\"{u}ckel-Su-Schrieffer-Heeger method.\\
I. Carbon-carbon bond lengths \\
in polycyclic aromatic hydrocarbons
}

\author{
\textbf{Jan H. Kwapisz$^a$ and Leszek Z. Stolarczyk$^{b*}$}
}
\date{\today}

\maketitle

\abstract{
  The equilibrium carbon-carbon bond lengths in $\pi$-electron hydrocarbons are very sensitive to the electronic ground-state characteristic. In the recent two papers by Stolarczyk and Krygowski (\textit{J. Phys. Org. Chem.}, this issue) a simple quantum approach, the Augmented H\"{u}ckel Molecular Orbital (AugHMO) model, is proposed for the qualitative, as well as quantitative, study of this phenomenon. The simplest realization of the AugHMO model is the H\"{u}ckel-Su-Schrieffer-Heeger (HSSH) method, in which the resonance integral $\beta$ of the HMO model is a linear function the bond length. 

  In the present paper the HSSH method is applied in a study of carbon-carbon bond lengths in a set of 34 selected policyclic aromatic hydrocarbons (PAHs). This is exactly the set of molecules analyzed by Rieger and M\"{u}llen (\textit{J. Phys. Org. Chem.} \textbf{2010}, \textit{23}, 315) in the context of their electronic-excitation spectra. These PAHs have been obtained by chemical synthesis, but in most cases no diffraction data (by X-rays or neutrons) of sufficient quality is available to provide us with their geometry. On the other hand, these PAHs are rather big (up to 96 carbon atoms), and \textit{ab initio} methods of quantum chemistry are too expensive for reliable geometry optimization. That makes the HSSH method a very attractive alternative. 
  
  Our HSSH calculations uncover a modular architecture of certain classes of PAHs. For the studied molecules (and their fragments -- modules) we calculate the values of the aromaticity index HOMA.
} % 233 words

\vfill

\noindent $^*$ \emph{Correspondence to: Leszek Z. Stolarczyk, Faculty of Chemistry, University of Warsaw, Pasteura 1, PL-02-093 Warsaw, Poland.} \\
\emph{E-mail: leszek@tiger.chem.uw.edu.pl} \\[2ex]
a \emph{Faculty of Physics, University of Warsaw, Pasteura 5, PL-02-093 Warsaw, Poland} \\
b \emph{Faculty of Chemistry, University of Warsaw, Pasteura 1, PL-02-093 Warsaw, Poland}

\newpage

\section{Introduction} \label{sec:intro}

  The variability of the (equilibrium) carbon-carbon (C-C) bond lengths in $\pi$-electron hydrocarbons and all-carbon molecules (fullerenes, nanotubes, and graphene) is a consequence of the coupling between the nuclear framework and the system of \emph{mobile} electrons that occupy the molecular orbitals of the $\pi$-symmetry. This phenomenon poses a challenge to both experiment and theory. 
  
  On the experimental side one faces difficulties in obtaining good-quality monocrystals. Moreover, the usual diffraction techniques (by X-rays or neutrons) suffer from the limitations of the theoretical model which translates the diffraction data into the molecular geometry. This model, in which spherical atoms are subject to uncorrelated thermal motions, neglects the changes of the electronic density due to chemical bonding (X-ray diffraction), as well as the rigid-body motions of the whole molecule. The latter effects, affecting both the X-ray and neutron studies, are especially important for large and rigid structures of $\pi$-electron hydrocarbons.
  
  On the theory side there are convergence problems of quantum-mechanical calculations with respect to the orbital basis sets and the electronic-correlation contributions. These problems become especially acute with the growing size of the studied molecules. As it has been already in the past, a carefully crafted semi-empirical quantum approach may come to the rescue.
  
  In the two recent papers by Stolarczyk and Krygowski~\cite{Stol2020a,Stol2020b} a very simple Augmented H\"{u}ckel Molecular Orbital (AugHMO) model has been designed for this purpose. It allows for a qualitative, as well as quantitative, study of the bond-length variation in general $\pi$-electron molecules (i.e., those which are planar or locally planar). The simplest realization of the AugHMO model, the the H\"{u}ckel-Su-Schrieffer-Heeger (HSSH) method, was parametrized~\cite{Stol2020b} for the $\pi$-electron hydrocarbons and all-carbon molecules. It was demonstrated~\cite{Stol2020b} that the HSSH method is capable of a very good description of the C-C bond lengths in a variety of $\pi$-electron hydrocarbons and carbon systems, including fullerene C$_{60}$, polyacetylene, and graphene.
  
  In the present paper we employ the HSSH method to calculate the equilibrium C-C bond lengths for a set of the polycyclic aromatic hydrocarbons (PAHs). These are precisely the molecules whose spectroscopic properties were summarized and analysed in a review paper by Riegel and M\"{u}llen~\cite{Riegel2010}. These PAHs have been obtained by chemical synthesis (many of them by M\"{u}llen and his coworkers), but in most cases no diffraction data of sufficient quality is available to provide us with their geometries. On the other hand, these PAHs are rather big (up to 96 C atoms), and \textit{ab initio} methods of quantum chemistry are too expensive for a reliable geometry optimization.
  
  In our paper we keep the classification of PAHs, as well as the numbering of the molecules, employed in Ref.~\cite{Riegel2010}:
\begin{itemize}
\item Clarenes (i.e., all-benzenoid PAHs): molecules \textbf{1} -- \textbf{14}, see Sec.~\ref{sec:clar}, Figs.~\ref{fig:clar1} and \ref{fig:clar2}.
\item K-region PAHs: molecules \textbf{15} -- \textbf{20} (plus ovalene), see Sec.~\ref{sec:K-region}, Figs.~\ref{fig:PCO} and \ref{fig:KPAHs2}.
\item Phenacenes: molecules \textbf{21} -- \textbf{25}, see Sec.~\ref{sec:phen}, Fig.~\ref{fig:phen}.
\item Rylenes: molecules \textbf{26} -- \textbf{29}, see Sec.~\ref{sec:ryl}, Fig.~\ref{fig:ryl}.
\item Acenes: molecules \textbf{30} -- \textbf{34} (plus naphthalene), see Sec.~\ref{sec:ace}, Fig.~\ref{fig:ace}. 
\end{itemize}

\section{HSSH calculations} \label{sec:calcs}

  The HSSH bond-length optimization procedure is a (simplified) analog of the molecular-geometry optimization techniques that employ the analytical gradients and Hessians of the total molecular energy~\cite{Yama1994}. In the HSSH method the hydrocarbon under study is fully characterized by its topology of C-C bonds (which is coded in the form of the topological matrix~\cite{Stol2020a,Stol2020b}). 
  
  At the start of the calculations all the C-C bonds are put equal to the value for benzene, $R_{\rm ben}^{\rm e} = 1.397$ {\AA}. Then the usual H\"{u}ckel calculations follow, yielding the molecular orbitals (MOs) and the $\pi$-electron bond orders $p_{\rm bond}$. This is the 0th iteraction, fully equivalent to the standard H\"{u}ckel approach. In each subsequent step (iteration) the condition of the vanishing gradient of the total HSSH energy is enforced: this amounts to iterating the \emph{linear} bond-order bond-length (BO-BL) relationship:
\begin{equation} \label{eq:LBOBL}
R_{\rm bond}^{\rm e} = R^{\rm o} - x \, p_{\rm bond}^{\rm e} \,,
\end{equation}
where $R_{\rm bond}^{\rm e}$ and $p_{\rm bond}^{\rm e}$, respectively, are the equilibrium bond length and the $\pi$-electron bond order for a given C-C bond. $R^{\rm o} = 1.523$ {\AA} and $x = 0.189$ {\AA} are two geometrical parameters of the AugHMO model of $\pi$-electron hydrocarbons~\cite{Stol2020a,Stol2020b}. In the HSSH method for hydrocarbons~\cite{Stol2020b} there is also a third geometrical parameter $y = 0.2756$ {\AA} which determines the slope of a linear function $\beta (R)$ describing the dependence of the ,,resonance integral'' $\beta$ on the C-C bond length $R$. In order to improve the rate of convergence of the iterative procedure, the Hessian of the HSSH total energy is calculated and applied~\cite{Stol2020a}. In our calculations for PAHs we assumed that the iterations stopped when the subsequent values of bond-length differences dropped below $5 \cdot 10^{-6}$ {\AA|. To this end five ,,direct iterations'' using Eq.~(\ref{eq:LBOBL}) followed by two to three iterations involving the Hessian were always sufficient.

  Four points concerning the HSSH method~\cite{Stol2020a,Stol2020b} should be made clear. First:  this method optimizes \emph{only} the bond lengths, and it is assumed that the optimal valence angles (the C-C-C ones in the case of PAHs) are consistent with the optimized C-C bond lengths, and as close to $120^{\rm o}$ as possible. By reducing the molecular geometry to bond lengths only seems justified for such molecules as PAHs and contribute to the computational effectiveness of the HSSH method. Second: while Eq.~(\ref{eq:LBOBL}) looks like a relationship of \emph{local character}, it should be remembered that the $\pi$-electron bond orders $p_{\rm bond}$ are derived from the set of completely delocalized occupied $\pi$ molecular orbitals~\cite{Stol2020a}. And in the HMO model (as well as in the AugHMO one) these bond orders appear to be sensitive probes of molecular topology. Third: PAHs belong to \emph{the class of alternant $\pi$-electron hydrocarbons}~\cite{Stol2020b}, and within the HMO (AugHMO) model the Coulson-Rushbrooke theorem~\cite{Couls1940} holds: (i) the $\pi$-electron orbital energies for the occupied and unoccupied states are placed symmetrically with respect to the value of the ,,Coulomb integral'' $\alpha$, (ii)  the net $\pi$-electron charges~\cite{Stol2020a} at the C atoms are equal to zero. Four: in the HSSH calculations for hydrocarbons we employ the H\"{u}ckel energy units: 
\begin{equation} \label{eq:Huckel_units}
\alpha = 0 \,, \quad |\beta| \equiv |\beta(R_{\rm ben}^{\rm e})| = 1 \,,
\end{equation}
where $R_{\rm ben}^{\rm e} = 1.397$ {\AA} serves as the reference C-C bond length~\cite{Stol2020b}.Thus, the only empirical parameters of the HSSH model are the above-mentioned three geometrical parameters: $R^{\rm o}$, $x$, and $y$, plus $R_{\rm ben}$.

  We used our HSSH optimized bond lengths to calculate the values of the aromaticity index HOMA of Krygowski and coworkers~\cite{Krusz1972, Kryg1993, Kryg1996, Cyra1999}. HOMA (Harmonic-Oscillator Measure of Aromaticity) is currently considered the most important indicator of the aromatic character of $\pi$-electron molecules (or their fragments), based solely on the values of bond lengths. For a $\pi$-electron hydrocarbon the definition of HOMA reads as~\cite{Krusz1972, Kryg1993}
\begin{equation} \label{eq:HOMA1}
{\rm HOMA} = 1 - \frac{\alpha}{N} \sum_{\rm bond} (R_{\rm bond}^{\rm e} - R_{\rm opt})^2  \,,
\end{equation}  
where the summation goes over all $N$ carbon-carbon bonds in the $\pi$-electron hydrocarbon (or its fragment), $\alpha = 257.7$ {\AA}$^{-2}$ is a normalization constant, while $R_{\rm opt} = 1.388$ {\AA} represents the optimal value of the C-C bond length in the HOMA model. Although of energetic provenience~\cite{Krusz1972}, HOMA of Eq.~(\ref{eq:HOMA1}) is dimensionless and fulfils condition $0 \leq {\rm HOMA} \leq 1$ (smaller HOMA indicates lower aromatic character). It was found~\cite{Kryg1996, Cyra1999} that definition~(\ref{eq:HOMA1}) can be rewritten in a more revealing form: 
\begin{equation} \label{eq:HOMA2}
{\rm HOMA} = 1 - {\rm EN} - {\rm GEO} = 1 - \alpha (R_{\rm av} - R_{\rm opt})^2  - \frac{\alpha}{N} \sum_{\rm bond} (R_{\rm bond}^{\rm e} - R_{\rm av})^2  \,,
\end{equation}
where
\begin{equation} \label{eq:Rav}
R_{\rm av} = \frac{1}{N} \sum_{\rm bond} R_{\rm bond}^{\rm e} 
\end{equation}
is the average C-C bond length in the $\pi$-electron hydrocarbon (or its fragment). The components EN and GEO may be interpteted as some ,,dearomatization'' contributions originating from the bond-length elongation and bond alternation, respectively~\cite{Cyra1999}. Let us remark that the EN component is simply a quadratic function of $R_{\rm av}$, and that the EN and GEO components (and thus HOMA) are sensitive to the rounding errors in the values of $R_{\rm bond}^{\rm e}$.

\section{Modular architecture of PAHs} \label{sec:modarch}

  The carbon skeletons of PAHs are finite jigsaw cuts from the honeycomb lattice of graphene. At the perimeter of a PAH there is a number of methine groups ($>$CH) corresponding to the tertiary carbon atoms (each with two C neighbors); the remaining C atoms are of the quaternary kind (with three C neighbors). The varied shapes of PAHs determine the corresponding topologies of the C-C bonds, which may be represented by the so-called H\"{u}ckel graphs (see Ref.~\cite{Trinajstic:1992}, p. 28). we shall use such graphs in the figures throughout the paper: the C atoms are graph vertices, and the C-C $\sigma$ bonds are graph edges (the H atoms and the C-H bonds are absent). In the AugHMO model~\cite{Stol2020a,Stol2020b}, the topologies of the C-C bonds directly translate into the bond-length pattern. Our HSSH calculations for PAHs reveal a considerable spread (ca. $0.1$ {\AA}) of the equilibrium C-C bond lengths: from $1.363$ {\AA}  [bond a in pyrene (\textbf{15}), see Fig.~\ref{fig:PCO}], to $1.461$ {\AA} [bond m' in perylene (\textbf{26}), see Fig.~\ref{fig:ryl}]. These values are to be compared with the HSSH result for graphene~\cite{Stol2020b} equal to $1.424$ {\AA}. 
  
  Our HSSH calculations indicate that some classes of PAHs (clarenes, rylenes, and a subclass of the K-region PAHs) are well represented by a \emph{modular architecture}. The ground-state geometry of such a PAH can be assemblied by using certain standard molecular fragments (modules) of fixed geometries, connected by C-C bonds (linkers) of fixed lengths. The set of modules employed in the present study is depicted, in the form of H\"{u}ckel graphs, in Fig.~\ref{fig:ABCmodules}. These modules are assumed to have ,,natural'' symmetries, independent of the particular molecular environment. More details will be provided in sections devoted to clarenes (Sec.~\ref{sec:clar}), K-region PAHs (2) (Subsec.~\ref{ssec:KPAHs2}), and rylenes (Sec.~\ref{sec:ryl}).
  
\FloatBarrier
\begin{figure}[!h]
\begin{minipage}{\linewidth}
\centering
\includegraphics[scale=0.15]{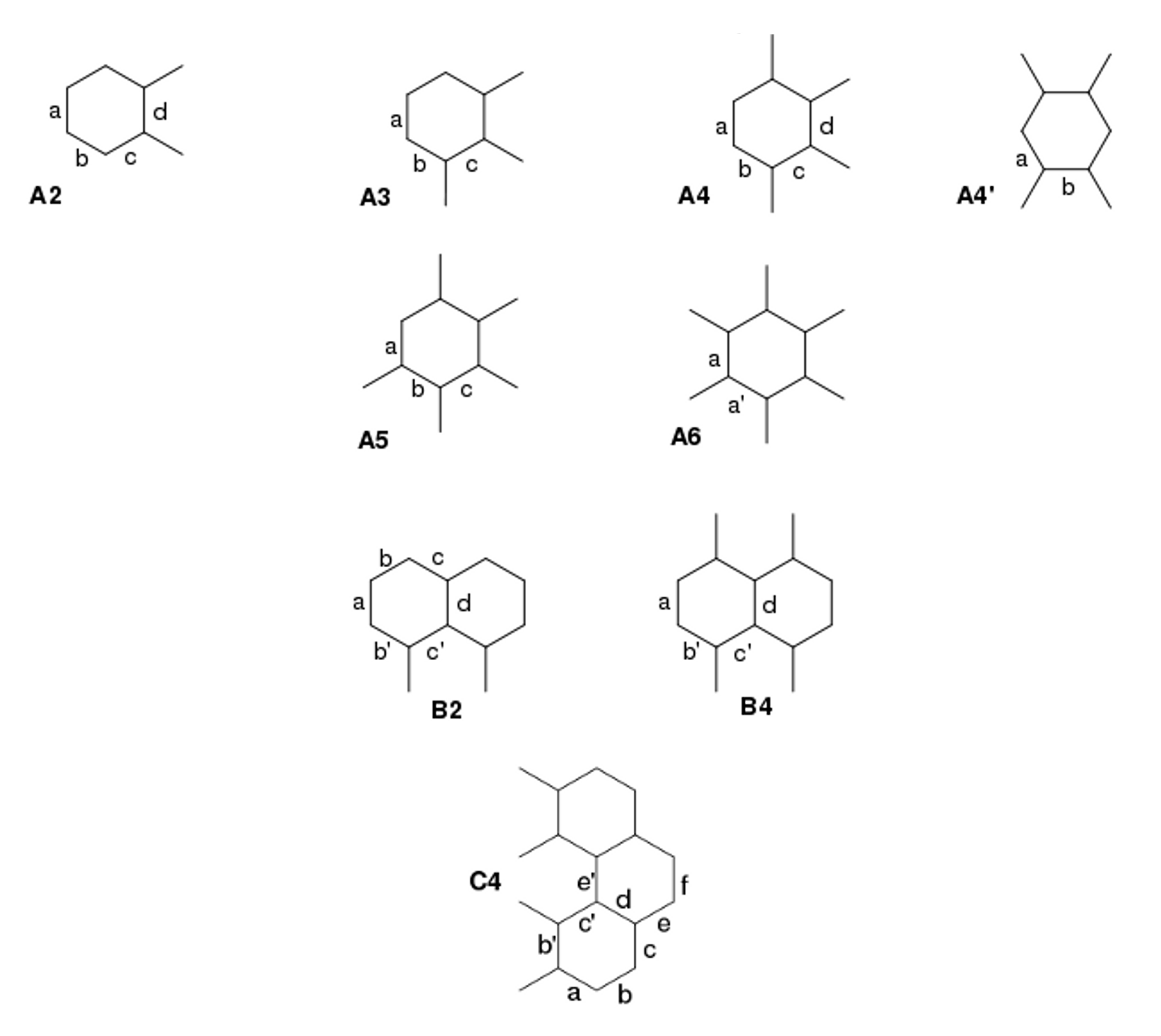}
\caption{A-, B-, and C-modules. H\"{u}ckel graphs, with symbols of symmetry nonequivalent bonds; directions of linkers (with no endpoint C-atoms) are indicated.}
\label{fig:ABCmodules}
\end{minipage}
\end{figure}
\FloatBarrier

  How the bond lengths corresponding to the modules and linkers were computed? Take the example of bond a in module A3: the length $\overline{R^{\rm e}_{\rm a}}$ was calculated by averaging over all the 28 symmetry-unrelated occurences of this type of bond in the molecules containig the A3-module (see Figs.~\ref{fig:clar1}, \ref{fig:clar2}, and \ref{fig:KPAHs2}). As a measure of the spread of the actual values of this bond length we used the absolute value of maximal difference,
\begin{equation} \label{eq:Delta}
\Delta_a = \textrm{max} | R_{\rm a}^{\rm e} - \overline{R_{\rm a}^{\rm e}} | \,,
\end{equation}
corresponding to the above-mentioned set of 28 values of $R_{\rm a}^{\rm e}$. 

\section{Clarenes} \label{sec:clar}

  Eric Clar proposed~\cite{Clar1972, Sola2013} that the all-benzenoid PAHs are special because of being ideal superpositions of six-electron units -- the ,,Clar sextets.'' Thus, it seems to be quite proper to call this class of PAHs ,,clarenes.''  The family of clarenes, as taken from the paper by Riegel and M\"{u}llen\cite{Riegel2010}, consists of 14 molecules, form convenience partitioned into two subsets shown in Figs.~\ref{fig:clar1} and \ref{fig:clar2}.
  
\FloatBarrier
\begin{figure}[!h]
\centering
\includegraphics[scale=0.15]{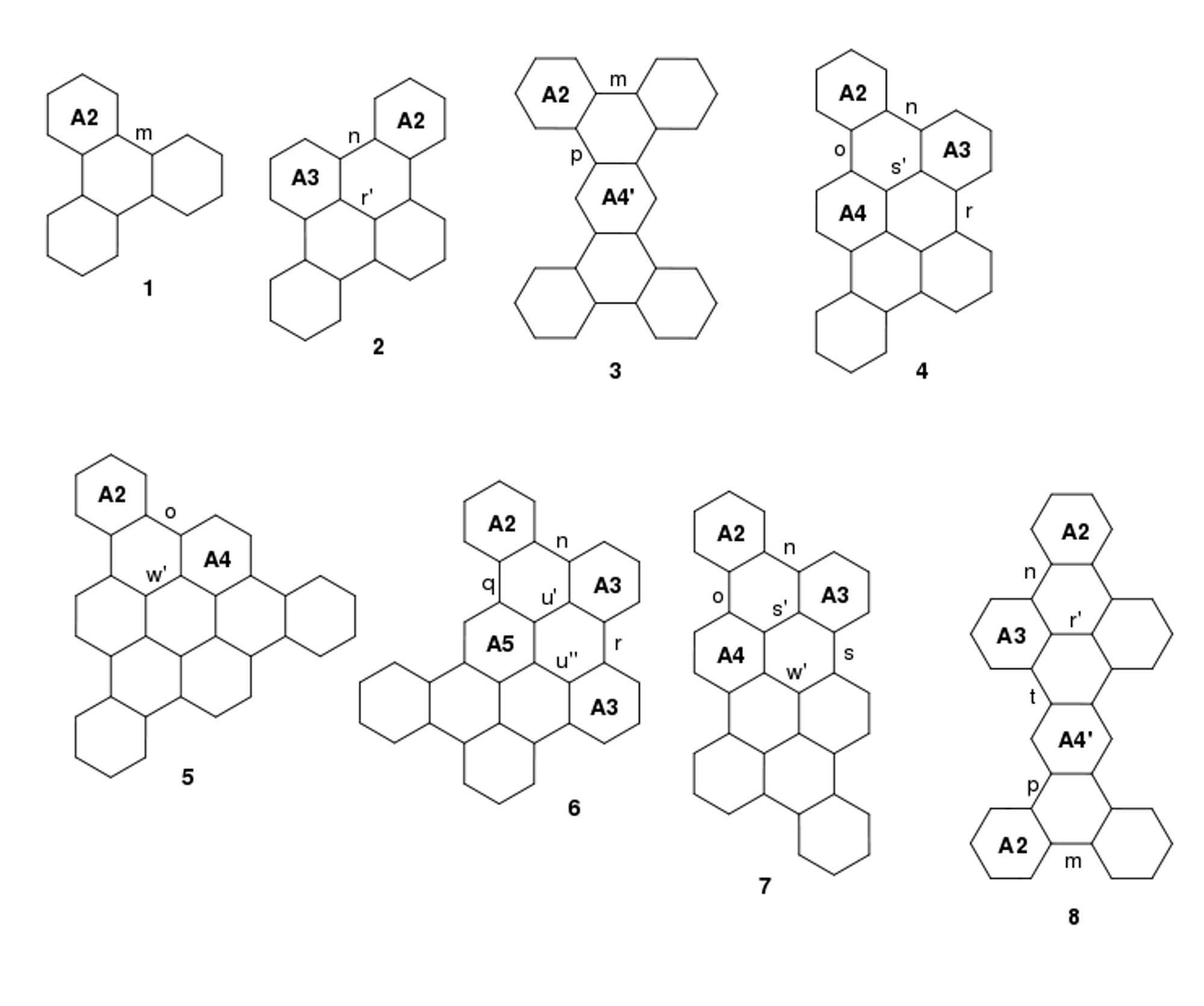}
\caption{Clarenes (1). H\"{u}ckel graphs, with symbols of symmetry nonequivalent A-modules and AA-linkers.}
\label{fig:clar1}
\end{figure}

\begin{figure}[!h]
\centering
\includegraphics[scale=0.15]{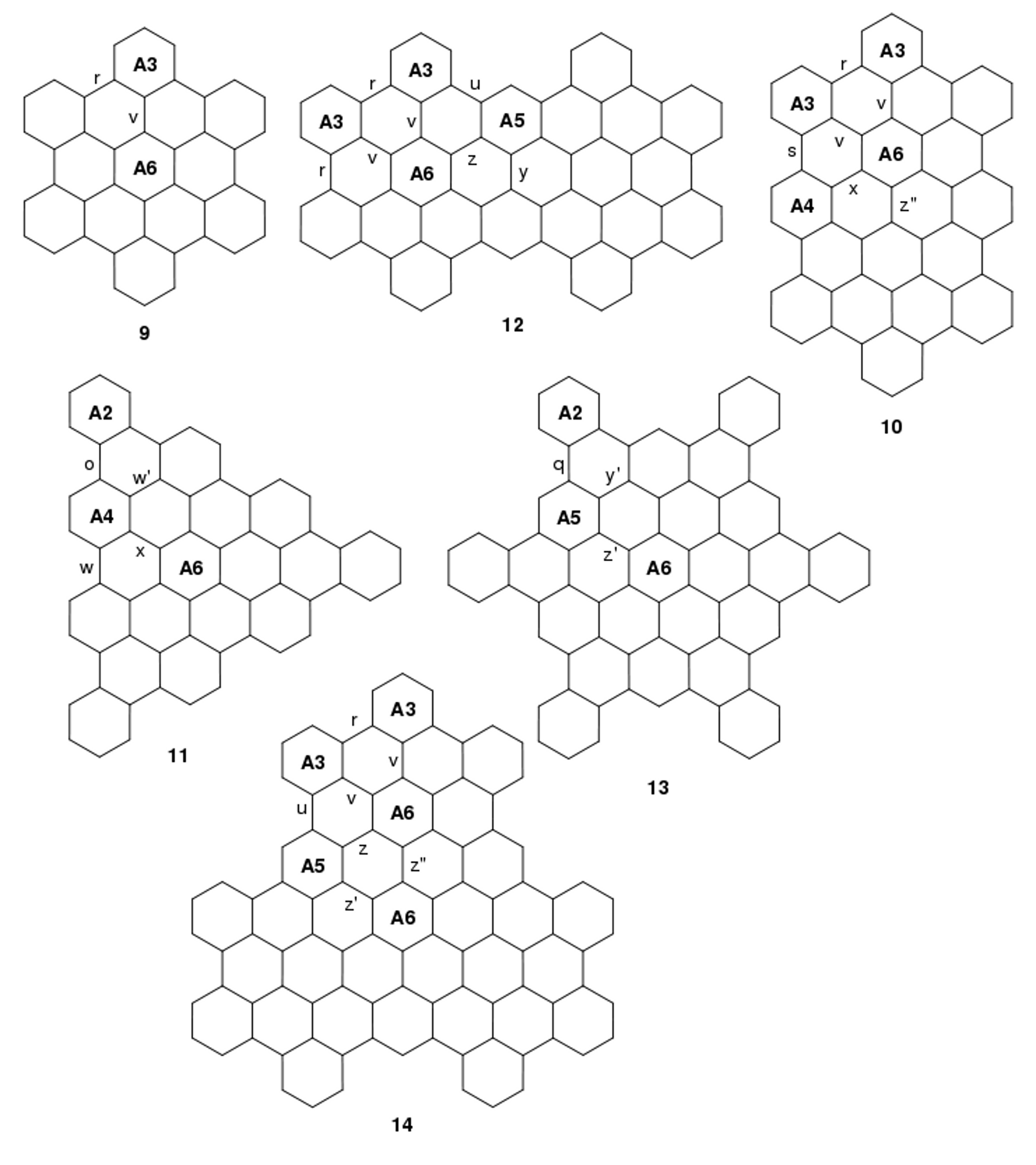}
\caption{Clarenes (2). H\"{u}ckel graphs, with symbols of symmetry nonequivalent A-modules and AA-linkers.}
\label{fig:clar2}
\end{figure}
\FloatBarrier

  Every molecule of clarene contain $6n$ carbon atoms and can be partitioned into six-carbon-atom benzene-like fragments, hereafter called the A-modules. Each A-module carries six $\pi$ electrons -- this is the above-mentioned Clar sextet. Within a given molecule the A-modules (depicted in Fig.~\ref{fig:ABCmodules}) are connected by C-C bonds, hereafter called the AA-linkers. The An-module (n = 2, 3, 4, 5 ,6) uses n linkers to connect with other modules; benzene may be considered a ,,honorary'' A0 module, while the phenyl group (not appearing in PAHs) is the A1-module. Modules A4 and A4' correspond to two diffrent arrangements of four linkers. Moreover, linkers connecting the same pair of modules (eg., A3-A5) may correspond to different linking topologies: see, e.g., linker u in molecule \textbf{12}, and linkers u' and u'' in molecule \textbf{6}. 
  
  Our HSSH calculations demonstrate that, with sufficient accuracy, each clarene can be additively assemblied from standard A-modules, connected via some standard AA-linkers. The results in Tabl.~\ref{tab:Re_A} and \ref{tab:Re_AAlink} are quite convincing: the An-modules (n = 2 -- 5) have well-defined geometries (with characteristic bond lengths), and the AA-linkers of a given type have nearly constant lengths. It is seen that within the A-modules the bond lengths are distictly shorter than those corresponding to the AA-linkers. Clearly, this is an indication that 
the Clar sextets are to some degree \emph{localized} within the clarene molecule. 

\FloatBarrier  
\begin{table}[!h] 
\centering
\begin{tabular}{|c||c|c|c|c|c|c|c|c|c|c|c|}
\hline & & \multicolumn{10}{c|}{Module} \\
\cline{3-12}  
     & \textbf{Benz.} & \multicolumn{2}{c|}{\textbf{A2}} & \multicolumn{2}{c|}{\textbf{A3}} & \multicolumn{2}{c|}{\textbf{A4}} & \multicolumn{2}{c|}{\textbf{A4'}} & \multicolumn{2}{c|}{\textbf{A5}} \\
  \hline & & & & & & & & & & & \\[-2.0ex]
Bond & $R^{\rm e}$ & $\overline{R^{\rm e}}$ & $\Delta$ & $\overline{R^{\rm e}}$ & $\Delta$ & $\overline{R^{\rm e}}$ & $\Delta$ & $\overline{R^{\rm e}}$ & $\Delta$ & $\overline{R^{\rm e}}$ & $\Delta$ \\
\hline
\hline a  & 1.397 & 1.406 & 0.002 & 1.397 & 0.003 & 1.388 & 0.002 & 1.402 & 0.000 & 1.403 & 0.000 \\
\hline b  &  ---  & 1.389 & 0.002 & 1.404 & 0.004 & 1.412 & 0.003 & 1.416 & 0.000 & 1.417 & 0.001 \\
\hline c  &  ---  & 1.409 & 0.002 & 1.415 & 0.002 & 1.410 & 0.001 &  ---  &  ---  & 1.416 & 0.002 \\
\hline d  &  ---  & 1.409 & 0.001 &  ---  &  ---  & 1.420 & 0.002 &  ---  &  ---  &  ---  &  ---  \\
\hline
\end{tabular}
\caption{A-modules (up to A5). HSSH equilibrium C-C bond lengths (in {\AA}).}
\label{tab:Re_A}
\end{table}

\begin{table}[!h] 
\begin{minipage}{\linewidth}
\centering
\begin{tabular}{|c|c|c|c||c|c|c|c|}
\hline \multicolumn{2}{c|}{} & & & \multicolumn{2}{c||}{} & & \\[-2.0ex]
 \multicolumn{2}{c|}{Linker} & $\overline{R^{\rm e}}$ & $\Delta$ & \multicolumn{2}{c||}{Linker} & $\overline{R^{\rm e}}$ & $\Delta$ \\
\hline \multicolumn{2}{l|}{\mbox{m}  (\textbf{A2-A2})} & 1.456 & 0.001 & \multicolumn{2}{l||}{\mbox{w } (\textbf{A4-A4})} & 1.447 &  ---  \\
\hline \multicolumn{2}{l|}{\mbox{n } (\textbf{A2-A3})} & 1.455 & 0.001 & \multicolumn{2}{l||}{\mbox{w'} (\textbf{A4-A4})} & 1.442 & 0.001 \\
\hline \multicolumn{2}{l|}{\mbox{o } (\textbf{A2-A4})} & 1.452 & 0.002 & \multicolumn{2}{l||}{\mbox{x } (\textbf{A4-A6})} & 1.439 &  ---  \\ 
\hline \multicolumn{2}{l|}{\mbox{p } (\textbf{A2-A4'})}& 1.456 & 0.000 & \multicolumn{4}{c|}{} \\
\hline \multicolumn{2}{l|}{\mbox{q } (\textbf{A2-A5})} & 1.454 & 0.001 & \multicolumn{4}{c|}{} \\
\hline
\hline \multicolumn{2}{l|}{\mbox{r } (\textbf{A3-A3})} & 1.455 & 0.001 & \multicolumn{2}{l||}{\mbox{y } (\textbf{A5-A5})} & 1.438 &  ---  \\ 
\hline \multicolumn{2}{l|}{\mbox{r'} (\textbf{A3-A3})} & 1.444 & 0.000 & \multicolumn{2}{l||}{\mbox{y'} (\textbf{A5-A5})} & 1.447 &  ---  \\
\hline \multicolumn{2}{l|}{\mbox{s } (\textbf{A3-A4})} & 1.452 & 0.001 & \multicolumn{2}{l||}{\mbox{z } (\textbf{A5-A6})} & 1.437 & 0.001 \\ 
\hline \multicolumn{2}{l|}{\mbox{s'} (\textbf{A3-A4})} & 1.443 & 0.001 & \multicolumn{2}{l||}{\mbox{z'} (\textbf{A5-A6})} & 1.436 & 0.000 \\ 
\hline \multicolumn{2}{l|}{\mbox{t } (\textbf{A3-A4'})}& 1.456 &  ---  & \multicolumn{4}{c|}{} \\
\hline \multicolumn{2}{l|}{\mbox{u } (\textbf{A3-A5})} & 1.453 & 0.000 & \multicolumn{4}{c|}{} \\
\hline \multicolumn{2}{l|}{\mbox{u'} (\textbf{A3-A5})} & 1.442 &  ---  & \multicolumn{4}{c|}{} \\
\hline \multicolumn{2}{l|}{\mbox{u"} (\textbf{A3-A5})} & 1.443 &  ---  & \multicolumn{4}{c|}{} \\
\hline \multicolumn{2}{l|}{\mbox{v } (\textbf{A3-A6})} & 1.440 & 0.003 & \multicolumn{2}{l||}{\mbox{z"} (\textbf{A6-A6})} & 1.436 & 0.002 \\ 
\hline
\end{tabular}
\caption{AA-linkers. HSSH equilibrium C-C bond lengths (in {\AA}).}
\label{tab:Re_AAlink}
\end{minipage}
\end{table}
\FloatBarrier

  The An-modules (n = 2 -- 5) necessarily have to be located at the perimeter of a given clarene. The interior of the molecule, if sufficiently big, is filled with the A6-modules. Quite surprisingly, A6 cannot be considered a rigid building block: in a less symmetric surrounding its perfect $6mm$ (C$_{6v}$) symmetry\footnote{We use here the symbols corresponding to the planar point groups} becomes visibly perturbed. The most important cases of A6 deformations are presented in Tabl.~\ref{tab:Re_A6}. In molecules \textbf{11} and \textbf{20} (see Fig.~\ref{fig:KPAHs2}) the molecular symmetry group $3m$ (C$_{3v}$) induces some \emph{bond alternation} in the central A6-module; however, this effect is absent in molecule \textbf{14} due to a different orientation of the molecular-symmetry elements with respect to the central A6 hexagon. A similar pattern of deformation is found in molecule \textbf{19}, despite its lower symmetry corresponding to the $m$ (C$_{\rm s}$) group. It is apparent that bond alternation is the softest mode of deformation for the A6-module. 

\FloatBarrier
\begin{table}[!h] 
\centering
\begin{tabular}{|c||c|c|c|c|}
\hline Molecule &$R_{\rm a'}^{\rm e}$ ({\AA}) &$R_{\rm a}^{\rm e}$ ({\AA}) &$\frac{1}{2}(R_{\rm a'}^{\rm e} + R_{\rm a}^{\rm e})$& $R_{\rm a'}^{\rm e} - R_{\rm a}^{\rm e}$ \\
\hline
\hline \textbf{11} & 1.420  & 1.414 & 1.417 & 0.006 \\
\hline \textbf{19} & 1.421  & 1.414 & 1.417 & 0.007 \\
\hline \textbf{20} & 1.424  & 1.413 & 1.419 & 0.010 \\
\hline
\end{tabular}
\caption{A6-modules, HSSH equilibrium C-C bond lengths (in {\AA}). In molecules \textbf{11}, \textbf{19}, and \textbf{20} a pattern of \emph{alternating} bonds (a' a)$_3$ is seen. Full averaging over the remaining molecules containing A6-modules gives $R_{\rm a'}^{\rm e} = R_{\rm a}^{\rm e} = 1.417$, $\Delta = 0.003$.}
\label{tab:Re_A6}
\end{table}
\FloatBarrier
  
  While it seems proven that the perimeter of a clarene assumes a definite geometric structure corresponding to some arrangement od the An-modules (n = 2 -- 5), its interior is expected to approach the structure of graphene (however, see the discussion in the last Section). To this end the equilibrium bond lengths corresponding to bonds a and a' in the A6 module (see Fig.~\ref{fig:ABCmodules}), and the equilibrium bond length corresponding to the A6-A6 linker (z'', see molecule \textbf{14} in  Fig.~\ref{fig:clar2}) should approach the HSSH value for graphene ($1.424$ {\AA}). For the largest PAH in this study, molecule \textbf{14} (96 C atoms), we found $R_{\rm a'}^{\rm e} = R_{\rm a}^{\rm e} = 1.419$ {\AA}, $R_{\rm z''}^{\rm e} = 1.435$ {\AA}, quite far from the graphene limit. However, when one averages over three bond lengths adjacent to any C atom in any A6 module, the result is very close to $1.424$ {\AA}. Thus, even in relatively small clarenes their interiors ,,prepare'' for becoming graphenelike.  It may be said with some exaggeration that a big clarene molecule has a ,,hard shell'' made of some An-modules (n = 2 -- 5), protecting an interior consisting of ,,softer'' A6-modules, which, with the growing size of the molecule, loose their identity and melt into the homogenous honeycomb lattice of graphene.
  
  The inspection of the HOMA values and its components in Tabl.~\ref{tab:homa_A} indicates that the An-modules retain, to some degree, the aromaticity of the benzene molecule. It is seen that it is EN (and thus $\mathrm{R_{\rm av}}$) which is mostly responsible for diminishing of HOMA with the increase of n (the number of linkers). The result for graphene suggests its low aromaticity, despite the fact that graphene is the most stable of all carbon and $\pi$-electron hydrocarbon molecules (as the molecular enthalpies of formation attest). 
 
\FloatBarrier
\begin{table}[!h] 
\centering
\begin{tabular}{|l||c|c|c|c|}
\hline Module & $\mathrm{R_{\rm av}}$ ({\AA}) & EN & GEO & HOMA \\
\hline
\hline \textbf{Benzene} & 1.397 & 0.021 & 0.000 & 0.979 \\
\hline \textbf{A2}      & 1.402 & 0.051 & 0.021 & 0.928 \\
\hline \textbf{A3}      & 1.405 & 0.076 & 0.013 & 0.911 \\
\hline \textbf{A4}      & 1.409 & 0.110 & 0.043 & 0.847 \\
\hline \textbf{A4'}     & 1.406 & 0.086 & 0.012 & 0.902 \\
\hline \textbf{A5}      & 1.412 & 0.150 & 0.011 & 0.839 \\
\hline \textbf{A6}      & 1.417 & 0.217 & 0.000 & 0.783 \\
\hline \textbf{Graphene}& 1.424 & 0.334 & 0.000 & 0.666 \\
\hline 
\end{tabular}
\caption{A-modules. $\mathrm{R_{\rm av}}$, EN, GEO, and HOMA values.}
\label{tab:homa_A}
\end{table}

  The molecular values of HOMA and its components collected in Tabl.~\ref{tab:homa_clar} show that the aromatic character of clarenes is lowering with their size. Both the EN and GEO components contribute to this effect. As clarenes grow larger, one finds that (i) the value of $\mathrm{R_{\rm av}}$ increases and thus departs from the HOMA value of $\mathrm{R_{\rm opt}}$, and (ii) the HOMO-LUMO gap (as calculated with the HSSH model) diminishes. Even if big clarenes are still far from the graphene limit, one may speculate that the metallic character and aromaticity are in conflict. It should be also remembered that in a PAH all the quaternary C atoms are completely inert~\cite{scott2015}, and thus all the chemistry takes place at the perimeter of the molecule, involving the methine groups. Therefore, the interior of PAHs (and clarenes in particular) may be of little (or no at all) relevance to their chemical reactivity.

\FloatBarrier
\begin{table}[!h] 
\centering
\begin{tabular}{|c||c|c|c|c|}
\hline Molecule & $\mathrm{R_{\rm av}}$ ({\AA}) & EN & GEO & HOMA \\
\hline
\hline \textbf{1}  & 1.409  & 0.119 & 0.106 & 0.775 \\
\hline \textbf{2}  & 1.412  & 0.147 & 0.105 & 0.748 \\
\hline \textbf{3}  & 1.412  & 0.143 & 0.115 & 0.742 \\
\hline \textbf{4}  & 1.413  & 0.164 & 0.103 & 0.733 \\
\hline \textbf{5}  & 1.414  & 0.175 & 0.099 & 0.727 \\
\hline \textbf{6}  & 1.414  & 0.176 & 0.102 & 0.722 \\
\hline \textbf{7}  & 1.414  & 0.175 & 0.100 & 0.725 \\
\hline \textbf{8}  & 1.413  & 0.158 & 0.112 & 0.730 \\
\hline \textbf{9}  & 1.416  & 0.199 & 0.090 & 0.711 \\
\hline \textbf{10} & 1.417  & 0.219 & 0.080 & 0.701 \\
\hline \textbf{11} & 1.416  & 0.208 & 0.086 & 0.706 \\
\hline \textbf{12} & 1.417  & 0.222 & 0.082 & 0.696 \\
\hline \textbf{13} & 1.417  & 0.214 & 0.086 & 0.700 \\
\hline \textbf{14} & 1.418  & 0.238 & 0.073 & 0.689 \\
\hline 
\end{tabular}
\caption{Clarenes. Molecular values of $\mathrm{R_{\rm av}}$ (in {\AA}), EN, GEO, and HOMA.}
\label{tab:homa_clar}
\end{table}

%\newpage

\section{K-region PAHs} \label{sec:K-region}

 Riegel and M\"{u}llen~\cite{Riegel2010} coined the name ,,K-region PAH'' to denote the PAH which can be derived from a clarene having one or more ,,bays'' within its perimeter. The ,,bay'' consist of four C atoms arranged as in the s-cis butadiene molecule (at the ,,mouth'' of the bay there are two methine groups, causing some steric tension). Examples of such bays are seen in clarenes \textbf{1} - \textbf{12} and \textbf{14}. The K-region emerges when the mouth of the bay is closed with a two-carbon insertion of the formula --CH--CH-- ; we shall call this insertion the ,,f-handle'', alluding to the bond f in phenanthrene (\textbf{21}), see Fig.~\ref{fig:phen}. Obviously, the new molecule does not fit the clarene paradigm, and the K-region (corresponding to the f-handle) introduces some olefinic character to the molecule (for a detailed discussion on the chemical and spectral properties of the K-region PAHs, see Ref.~\cite{Riegel2010})

 The K-region PAHs of Ref.~\cite{Riegel2010} may be conveniently divided into two groups: one containing pyrene and coronene (and ovalene, added here), and second composed of molecules which may be derived from ,,superbenzene'' (\textbf{9}), see Fig.~\ref{fig:clar2}. 

\subsection{K-region PAHs (1): pyrene, coronene, and ovalene}

  Pyrene, coronene, and ovalene, depicted in Fig.~\ref{fig:PCO}, are well-known PAHs. Their C-C bond lengths are listed in Tabl.~\ref{tab:Re_PCO}: a closer inspection of these data indicates that no obvious transferable fragments (modules) can be identified. This is a striking contrast with the clarenes of the previous Section. One can identify the ,,olefinic'' f-handles as the a-bonds in all three molecules, and the f-bond in ovalene.

\FloatBarrier
\begin{figure}[!h] 
\centering
\includegraphics[scale=0.15]{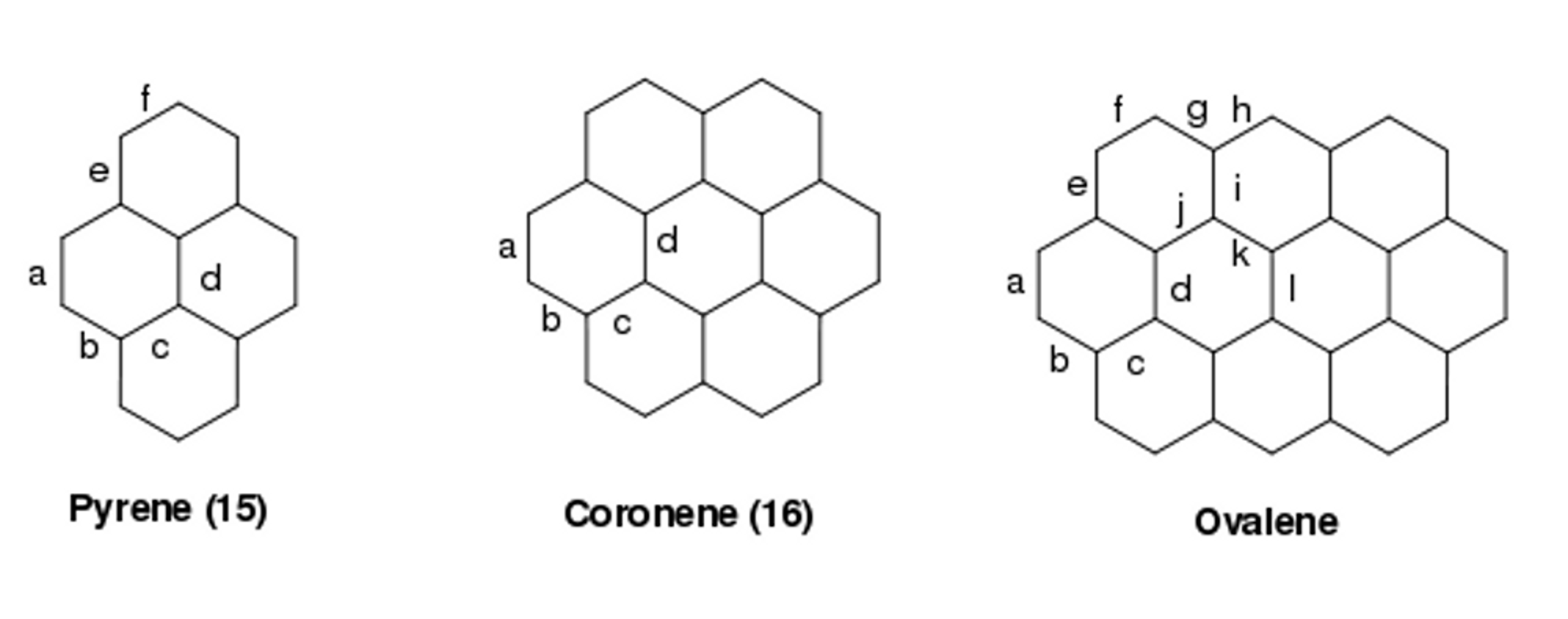}
\caption{K-region PAHs (1): pyrene, coronene, and ovalene. H\"{u}ckel graphs, with symbols of symmetry nonequivalent bonds.}
\label{fig:PCO}
\end{figure}
\FloatBarrier

\begin{table}[!h] 
\centering
\begin{tabular}{|c||c|c|c|}
\hline      & \multicolumn{3}{c|}{Molecule} \\
\hline
\hline Bond & \textbf{Pyrene} & \textbf{Coronene} & \textbf{Ovalene} \\
\hline a & 1.363 & 1.374 & 1.382 \\
\hline b & 1.444 & 1.429 & 1.419 \\
\hline c & 1.418 & 1.411 & 1.412 \\
\hline d & 1.432 & 1.430 & 1.425 \\
\hline e & 1.406 &  ---  & 1.438 \\
\hline f & 1.397 &  ---  & 1.367 \\
\hline g &  ---  &  ---  & 1.440 \\
\hline h &  ---  &  ---  & 1.405 \\
\hline i &  ---  &  ---  & 1.422 \\
\hline j &  ---  &  ---  & 1.432 \\
\hline k &  ---  &  ---  & 1.417 \\
\hline l &  ---  &  ---  & 1.438 \\
\hline 
\end{tabular}
\caption{Pyrene, coronene, and ovalene. HSSH equilibrium C-C bond lengths (in {\AA}).}
\label{tab:Re_PCO}
\end{table}
\FloatBarrier

 The molecular values of HOMA and its components are collected in Tabl.~\ref{tab:homa_PCO}. Although the general trend is similar to that for clarenes, Tabl.~\ref{tab:homa_clar}, this time also the GEO contribution is important. Let us note that HOMA for coronene is much lower than for benzene, and not very different from that of graphene, see Tabl.~\ref{tab:homa_A}.

\begin{table}[!h] 
\centering
\begin{tabular}{|c||c|c|c|c|}
\hline Molecule &$\mathrm{R_{\rm av}}$ ({\AA}) & EN & GEO & HOMA \\
\hline
\hline \textbf{Pyrene}   & 1.411  & 0.141 & 0.142 & 0.717 \\
\hline \textbf{Coronene} & 1.414  & 0.179 & 0.118 & 0.703 \\
\hline \textbf{Ovalene}  & 1.416  & 0.203 & 0.118 & 0.679 \\
\hline
\end{tabular}
\caption{Pyrene, coronene, and ovalene. Molecular values of $\mathrm{R_{\rm av}}$, EN, GEO, and HOMA.}
\label{tab:homa_PCO}
\end{table}
\FloatBarrier

%\newpage

\subsection{K-region PAHs (2): ,,superbenzene'' derivatives}\label{ssec:KPAHs2}

  The second group of the K-region PAHs of Ref.~\cite{Riegel2010} can be derived by inserting the f-handles into the bay regions of clarene \textbf{9}, called~\cite{Iyer1997} ,,superbenzene'' (according to that nomenclature, clarene \textbf{12} is ,,supernaphthalene'', and clarene \textbf{14} is ,,superphenalene''). One insertion gives molecule \textbf{17}, two (different) insertions -- molecules \textbf{18} and \textbf{19}, and three insertions -- molecule \textbf{20}, see Fig.~\ref{fig:KPAHs2}.
  
\FloatBarrier
\begin{figure}[!h] 
\centering
\includegraphics[scale=0.15]{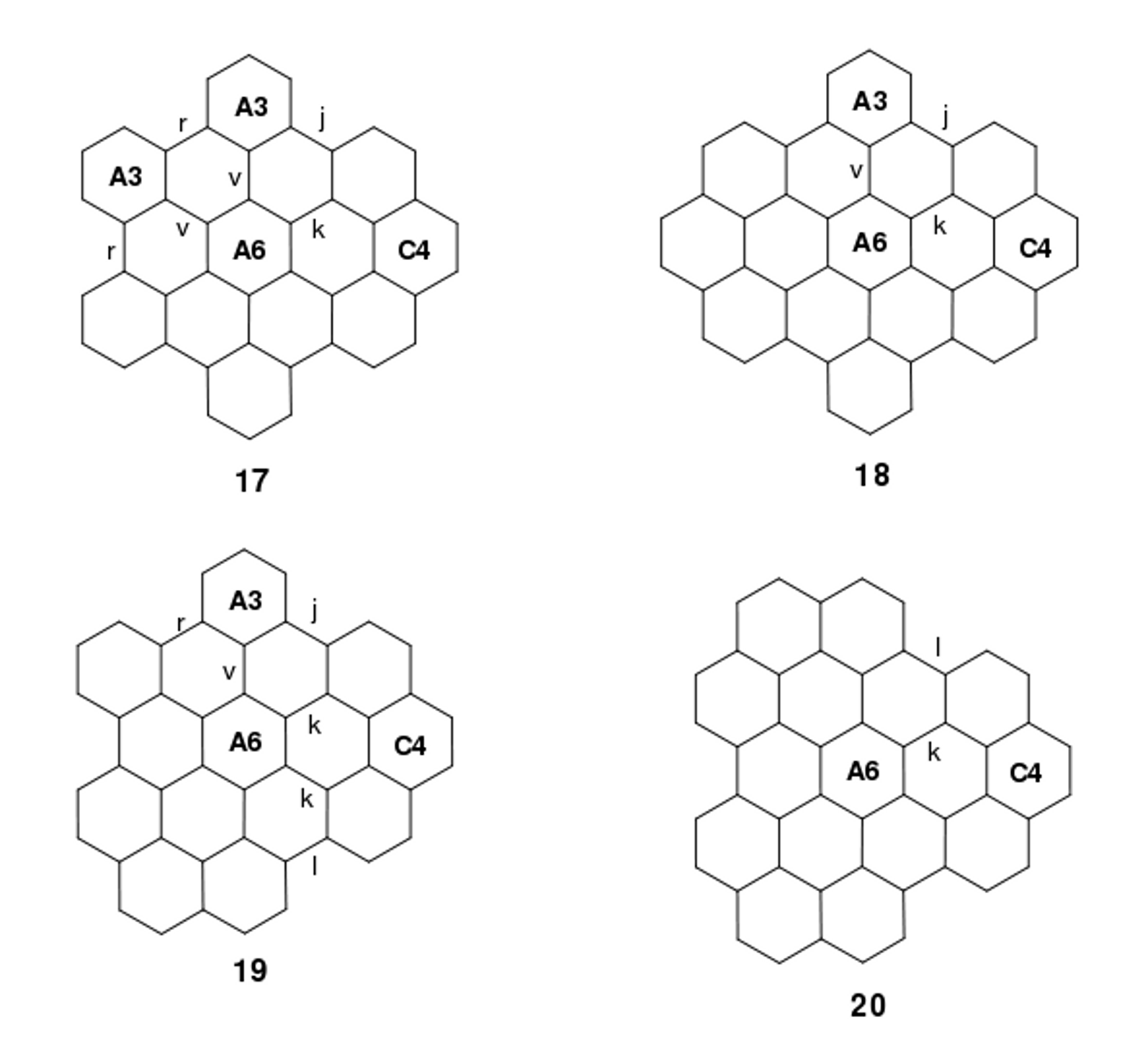}
\caption{K-region PAHs (2). H\"{u}ckel graphs, with symbols of symmetry nonequivalent A- and C-modules, and linkers.}
\label{fig:KPAHs2}
\end{figure}
\FloatBarrier

  In molecules (\textbf{17} - \textbf{20}) one may identify the A3 and A6 modules corresponding to clarene \textbf{9}, with practically ,,frozen'' structures, see Tabl.~\ref{tab:Re_AAlink} (also linkers r and v are unchanged). But a new module can be identified: this is the C4-module (see Fig.~\ref{fig:ABCmodules}), involving fourteen C atoms (and fourteen $\pi$ electrons), which resembles the phenanthrene molecule (\textbf{9}), the ,,C0-module''. As seen in Tabl.~\ref{tab:Re_C}, the (averaged) equilibrium C-C bond lengths of the C4-module are pretty well transferable, and so are the corresponding linkers (linker k being here a bit of exception). 

\FloatBarrier
\begin{table}[!h] 
\begin{minipage}{\linewidth}
\centering
\begin{tabular}{|c||c|c|c||c|c|c|c|}
\hline & \textbf{Phenanth.} & \multicolumn{2}{c||}{\textbf{C4}} & \multicolumn{4}{c|}{} \\
\hline & & & & \multicolumn{2}{c||}{} & & \\[-2.0ex]
      Bond & $R^{\rm e}$ & $\overline{R^{\rm e}}$ & $\Delta$ & \multicolumn{2}{c||}{Linker} & $\overline{R^{\rm e}}$ & $\Delta$ \\
\hline
\hline a  & 1.411 & 1.419 & 0.003  & \multicolumn{2}{l||}{\mbox{j } (\textbf{A3-C4})} & 1.449 & 0.001 \\
\hline b  & 1.385 & 1.381 & 0.002  & \multicolumn{2}{l||}{\mbox{k}  (\textbf{A6-C4})} & 1.436 & 0.005 \\						  
\hline b' & 1.385 & 1.408 & 0.001  & \multicolumn{2}{l||}{\mbox{l } (\textbf{C4-C4})} & 1.442 & 0.001 \\
\hline c  & 1.416 & 1.421 & 0.003  & \multicolumn{4}{c|}{} \\
\hline c' & 1.414 & 1.425 & 0.001  & \multicolumn{4}{c|}{} \\
\hline d  & 1.410 & 1.412 & 0.001  & \multicolumn{4}{c|}{} \\
\hline e  & 1.442 & 1.434 & 0.001  & \multicolumn{4}{c|}{} \\
\hline e' & 1.446 & 1.432 & 0.001  & \multicolumn{4}{c|}{} \\
\hline f  & 1.365 & 1.370 & 0.001  & \multicolumn{4}{c|}{} \\
\hline
\end{tabular}
\caption{C-modules and AC- and CC-linkers. HSSH equilibrium C-C bond lengths (in {\AA}).}
\label{tab:Re_C}
\end{minipage}
\end{table}
\FloatBarrier

The values of HOMA and its components for phenanthrene and the C4-module are collected in Tabl.~\ref{tab:homa_C}. The molecular values of HOMA and its components presented in Tabl.~\ref{tab:homa_KPAHs} show that each insertion of the f-handle into clarene lowers its HOMA and thus diminishes its aromatic character.

\FloatBarrier
\begin{table}[!h]
\centering
\begin{tabular}{|l||c|c|c|c|}
\hline Module & $\mathrm{R_{\rm av}}$ ({\AA}) & EN & GEO & HOMA \\
\hline
\hline \textbf{Phenanth.} & 1.408  & 0.107 & 0.129 & 0.764 \\
\hline \textbf{C4}        & 1.413  & 0.158 & 0.093 & 0.749 \\
\hline 
\end{tabular}
\caption{C-modules. $\mathrm{R_{\rm av}}$, EN, GEO, and HOMA values.}
 \label{tab:homa_C}
\end{table}

\begin{table}[!h]
\centering
\begin{tabular}{|c||c|c|c|c|}
\hline Molecule &$\mathrm{R_{av}}$ ({\AA}) & EN & GEO & HOMA \\
\hline
\hline \textbf{9}  & 1.416  & 0.199 & 0.090 & 0.711 \\
\hline \textbf{17} & 1.416  & 0.206 & 0.091 & 0.703 \\
\hline \textbf{18} & 1.417  & 0.213 & 0.091 & 0.696 \\
\hline \textbf{19} & 1.417  & 0.213 & 0.092 & 0.695 \\
\hline \textbf{20} & 1.418  & 0.225 & 0.113 & 0.662 \\
\hline
\end{tabular}
\caption{``Superbenzene'' (\textbf{9}) and K-region PAHs (2). Molecular values of $\mathrm{R_{\rm av}}$, EN, GEO, and HOMA.}
\label{tab:homa_KPAHs}
\end{table}
\FloatBarrier
 
%\newpage
 
\section{Phenacenes} \label{sec:phen}

  The first five members of the family of phenacenes are shown in Fig.~\ref{fig:phen}. With the growing size they approach a limit of a $\pi$-electron polymer (not the 2D structure of graphene). The bond lengths of phenacenes are presented in Tabl.~\ref{tab:Re_phen}: here also we see no clue to identify transferable modules, although some regularities in bond lengths are clearly visible.

\FloatBarrier
\begin{figure}[!h]
\centering
\includegraphics[scale=0.15]{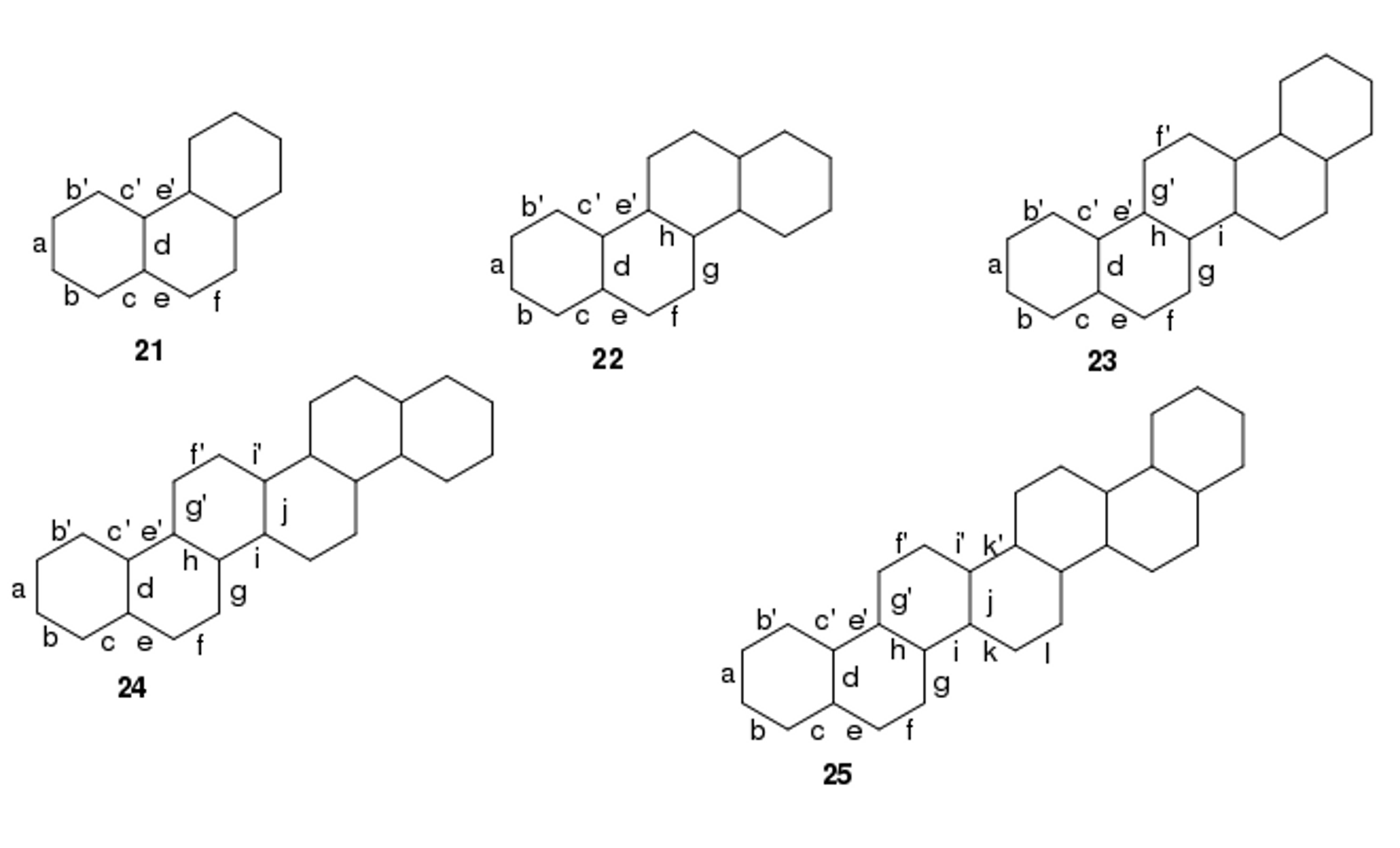}
\caption{Phenacenes. H\"{u}ckel graphs, with symbols of symmetry nonequivalent bonds.}
\label{fig:phen}
\end{figure}
\FloatBarrier

\begin{table}[!h] 
\centering
\begin{tabular}{|c||c|c|c|c|c|}
\hline      & \multicolumn{5}{c|}{Molecule} \\
\hline Bond &\textbf{21} &\textbf{22} &\textbf{23} &\textbf{24} &\textbf{25}  \\
\hline
\hline a & 1.411 & 1.415 & 1.413 & 1.414 & 1.414 \\
\hline b & 1.385 & 1.382 & 1.383 & 1.382 & 1.383 \\
\hline b'& 1.385 & 1.382 & 1.383 & 1.383 & 1.383 \\
\hline c & 1.416 & 1.419 & 1.418 & 1.418 & 1.418 \\
\hline c'& 1.414 & 1.418 & 1.416 & 1.417 & 1.417 \\
\hline d & 1.410 & 1.411 & 1.411 & 1.411 & 1.411 \\
\hline e & 1.442 & 1.435 & 1.438 & 1.437 & 1.437 \\
\hline e'& 1.446 & 1.439 & 1.442 & 1.441 & 1.441 \\
\hline f & 1.365 & 1.370 & 1.368 & 1.369 & 1.368 \\
\hline f'&  ---  &  ---  & 1.376 & 1.374 & 1.375 \\
\hline g &  ---  & 1.433 & 1.436 & 1.435 & 1.435 \\
\hline g'&  ---  &  ---  & 1.426 & 1.428 & 1.427 \\
\hline h &  ---  & 1.400 & 1.404 & 1.403 & 1.403 \\
\hline i &  ---  &  ---  & 1.431 & 1.434 & 1.433 \\
\hline i'&  ---  &  ---  &  ---  & 1.429 & 1.428 \\
\hline j &  ---  &  ---  &  ---  & 1.407 & 1.406 \\
\hline k &  ---  &  ---  &  ---  &  ---  & 1.431 \\
\hline k'&  ---  &  ---  &  ---  &  ---  & 1.437 \\
\hline l &  ---  &  ---  &  ---  &  ---  & 1.372 \\
\hline
\end{tabular}
\caption{Phenacenes. HSSH equilibrium C-C bond lengths (in {\AA}).}
\label{tab:Re_phen}
\end{table}

 Our HSSH equilibrium C-C bond lengths in phenacenes may be compared to those calculated by Firouzi and Zahedi~\cite{Firouzi:2008} by means of the \emph{ab initio} B3LYP/6-31G(d) approach. The agreement is satisfactory, with the exception of the ,,bridge bonds'' (of the symbols d, h, and j), for which the HSSH values are smaller by ca. $0.02$ {\AA} (see also acenes, Sec.~\ref{sec:ace}).

 The molecular values of HOMA and its components for phenacenes are presented in Tabl.~\ref{tab:homa_phen}. It seems that these quantities may converge to some limiting values in the case of the polyphenacene polymer.

%\FloatBarrier
\begin{table}[!h] 
\centering
\begin{tabular}{|c||c|c|c|c|}
\hline Molecule & $\mathrm{R_{\rm av}}$ ({\AA}) & EN & GEO & HOMA \\
\hline
\hline \textbf{21}  & 1.408  & 0.107 & 0.129 & 0.764 \\
\hline \textbf{22}  & 1.410  & 0.124 & 0.133 & 0.743 \\
\hline \textbf{23}  & 1.411  & 0.135 & 0.136 & 0.729 \\
\hline \textbf{24}  & 1.411  & 0.142 & 0.136 & 0.722 \\
\hline \textbf{25}  & 1.412  & 0.147 & 0.137 & 0.716 \\
\hline 
\end{tabular}
\caption{Phenacenes. Molecular values of $\mathrm{R_{\rm av}}$, EN, GEO, and HOMA.}
\label{tab:homa_phen}
\end{table}

\newpage

\section{Rylenes} \label{sec:ryl}

  in Fig.~\ref{fig:ryl} we present the first four members of the family of rylenes. As in the case of the family of phenacenes, with the growing size rylenes approach a limit of a $\pi$-electron polymer. However, in  difference to phenacenes, rylenes display modular architecture, based on the ten C atom, ten $\pi$-electron B2 and B4 modules, see Fig.~\ref{fig:ABCmodules}. They may be considered certain analogs of naphthalene (the ,,B0 module'', see Fig.~\ref{fig:ace}). The structures of the B2 and B4 modules are presented in Tabl.~\ref{tab:Re_B}, together with the lengths of the corresponding BB-linkers.
  
\FloatBarrier
\begin{figure}[!h]
\centering
\includegraphics[scale=0.15]{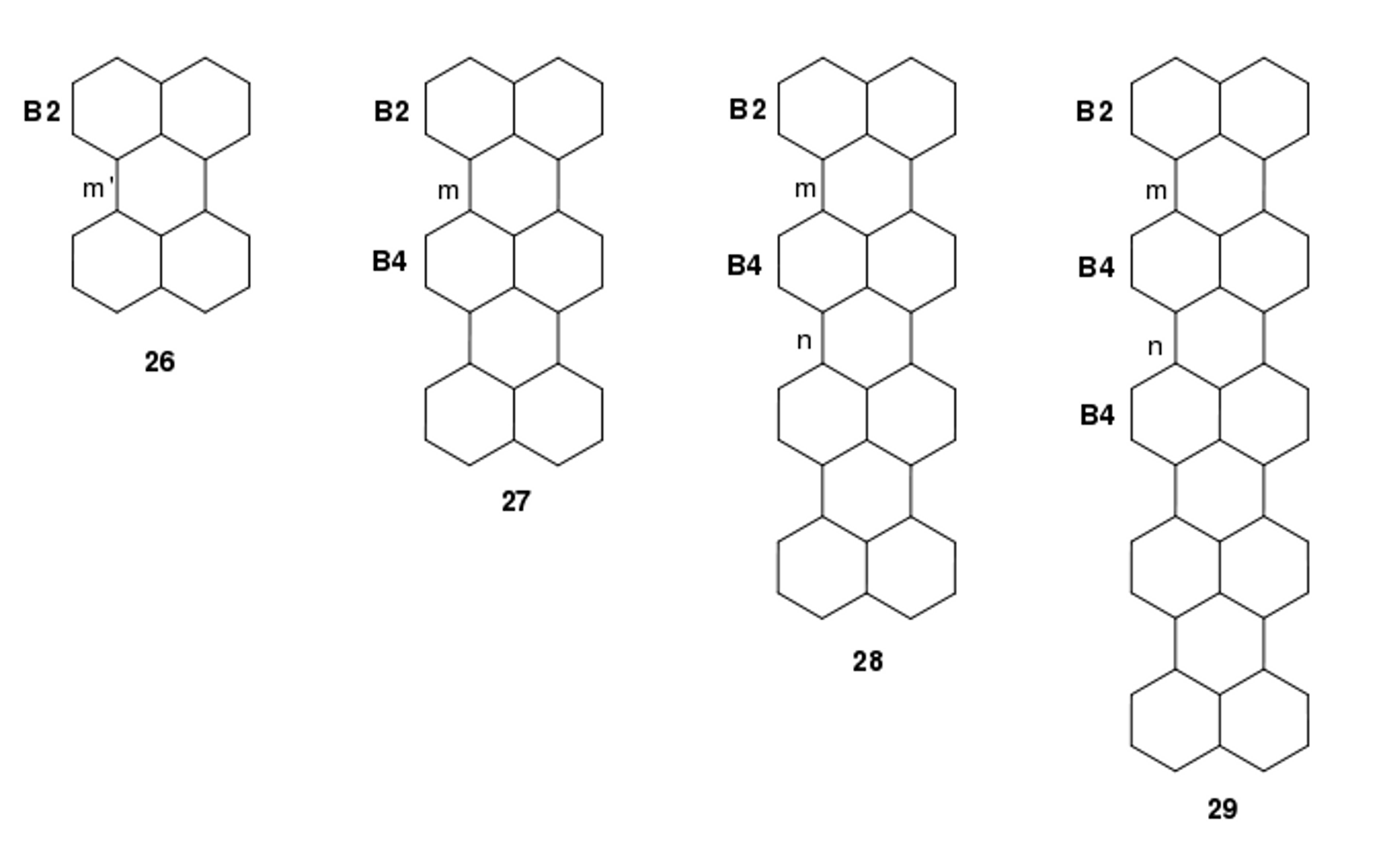}
\caption{Rylenes. H\"{u}ckel graphs, with symbols of symmetry nonequivalent B-modules and BB-linkers.}
\label{fig:ryl}
\end{figure}
\FloatBarrier

  The values of HOMA and its components for the B-modules are presented in Tabl.~\ref{tab:Re_B}, and the molecular values of these parameters calculated for rylenes appear in Tabl.~\ref{tab:homa_B}.

\begin{table}[!h] 
\centering
\begin{tabular}{|c||c|c|c|c|c||c|c|c|c|}
\hline & \textbf{Naphth.} & \multicolumn{2}{c|}{\textbf{B2}} & \multicolumn{2}{c||}{\textbf{B4}} & \multicolumn{4}{c|}{} \\
\hline & & & & & & \multicolumn{2}{c||}{} & & \\[-2.0ex]
  Bond & $R^{\rm e}$ & $\overline{R^{\rm e}}$ & $\Delta$ & $\overline{R^{\rm e}}$ & $\Delta$  & \multicolumn{2}{c||}{Linker} & $\overline{R^{\rm e}}$ & $\Delta$ \\
\hline
\hline a  & 1.421 & 1.414 & 0.001 & 1.405 & 0.002 & \multicolumn{2}{l||}{\mbox{m'}  (\textbf{B2-B2})} & 1.461 &  ---  \\
\hline b  & 1.377 & 1.381 & 0.001 &  ---  &  ---  & \multicolumn{2}{l||}{\mbox{m }  (\textbf{B2-B4})} & 1.458 & 0.001 \\
\hline b' &  ---  & 1.390 & 0.002 & 1.395 & 0.002 & \multicolumn{2}{l||}{\mbox{n    } (\textbf{B4-B4})} & 1.454 & 0.001 \\
\hline c  & 1.425 & 1.424 & 0.000 &  ---  &  ---  & \multicolumn{4}{c|}{} \\
\hline c' &  ---  & 1.427 & 0.000 & 1.426 & 0.000 & \multicolumn{4}{c|}{} \\
\hline d  & 1.414 & 1.415 & 0.000 & 1.417 & 0.001 & \multicolumn{4}{c|}{} \\
\hline
\end{tabular}
\caption{B-modules and BB-linkers. HSSH equilibrium C-C bond lengths (in {\AA}).}
\label{tab:Re_B}
\end{table}

\begin{table}[!h]
\centering
\begin{tabular}{|l||c|c|c|c|}
\hline Module & $\mathrm{R_{\rm av}}$ ({\AA}) & EN & GEO & HOMA \\
\hline
\hline \textbf{Naphth.}  & 1.406  & 0.081 & 0.125 & 0.794 \\
\hline \textbf{B2}       & 1.408  & 0.102 & 0.081 & 0.817 \\
\hline \textbf{B4}       & 1.410  & 0.126 & 0.045 & 0.829 \\
\hline 
\end{tabular}
\caption{B-modules. $\mathrm{R_{\rm av}}$, EN, GEO, and HOMA values.}
 \label{tab:homa_B}
\end{table}

\begin{table}[!h] 
\centering
\begin{tabular}{|c||c|c|c|c|}
\hline Molecule & $\mathrm{R_{\rm av}}$ ({\AA}) & EN & GEO & HOMA \\
\hline
\hline \textbf{26}  & 1.408  & 0.107 & 0.129 & 0.764 \\
\hline \textbf{27}  & 1.410  & 0.124 & 0.133 & 0.743 \\
\hline \textbf{28}  & 1.411  & 0.135 & 0.136 & 0.729 \\
\hline \textbf{29}  & 1.411  & 0.142 & 0.136 & 0.722 \\
\hline 
\end{tabular}
\caption{Rylenes. Molecular values of $\mathrm{R_{\rm av}}$, EN, GEO, and HOMA.}
\label{tab:homa_ryl}
\end{table}

\newpage

\section{Acenes} \label{sec:ace}

  The first six members of the family of acenes are shown in Fig.~\ref{fig:ace}. Acenes are similar to phenacenes of Sec.~\ref{sec:phen}. They also, with the growing size, approach a limit of a $\pi$-electron polymer. The bond lengths of acenes are presented in Tabl.~\ref{tab:Re_ace}: here also we see no clue to identify transferable modules despite some regularities in bond lengths.

\begin{figure}[!h]
\centering
\includegraphics[scale=0.15]{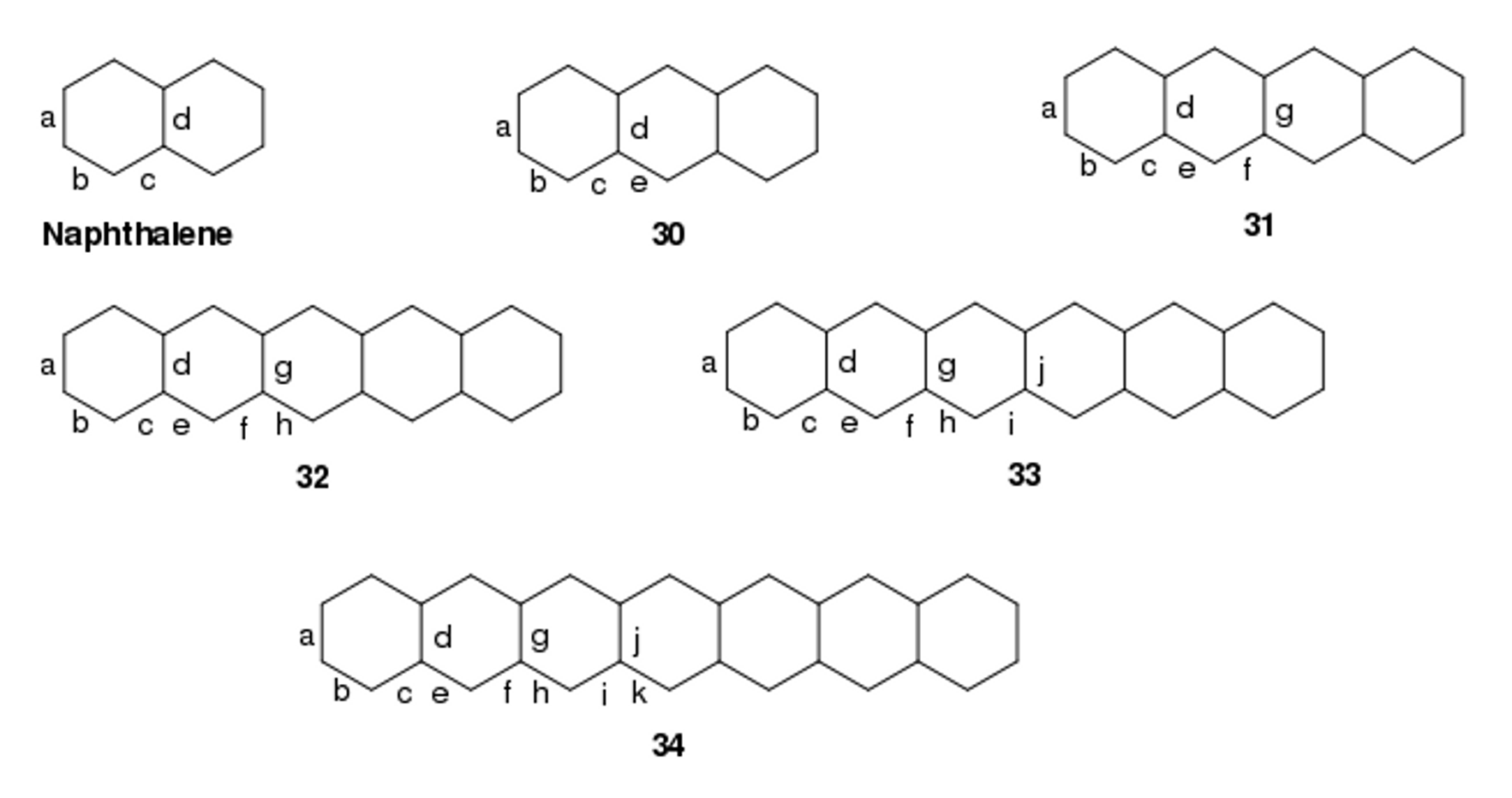}
\caption{Acenes. H\"{u}ckel graphs, with symbols of symmetry nonequivalent bonds.}
 \label{fig:ace}
\end{figure}

%\FloatBarrier
\begin{table}[!h] 
\centering
\begin{tabular}{|c||c|c|c|c|c|c|c|}
\hline      & \multicolumn{7}{c|}{Molecule} \\
\hline Bond & \textbf{Benzene} & \textbf{Naphth.} & \textbf{30} &\textbf{31} &\textbf{32} &\textbf{33} &\textbf{34}  \\
\hline
\hline a & 1.397 & 1.421 & 1.431 & 1.435 & 1.438 & 1.439 & 1.440 \\
\hline b &  ---  & 1.377 & 1.370 & 1.367 & 1.366 & 1.365 & 1.364 \\
\hline c &  ---  & 1.425 & 1.435 & 1.440 & 1.442 & 1.443 & 1.444 \\
\hline d &  ---  & 1.414 & 1.426 & 1.433 & 1.437 & 1.439 & 1.441 \\
\hline e &  ---  &  ---  & 1.404 & 1.395 & 1.390 & 1.388 & 1.387 \\
\hline f &  ---  &  ---  &  ---  & 1.416 & 1.422 & 1.425 & 1.427 \\
\hline g &  ---  &  ---  &  ---  & 1.432 & 1.437 & 1.439 & 1.441 \\
\hline h &  ---  &  ---  &  ---  &  ---  & 1.406 & 1.401 & 1.398 \\
\hline i &  ---  &  ---  &  ---  &  ---  &  ---  & 1.413 & 1.416 \\
\hline j &  ---  &  ---  &  ---  &  ---  &  ---  & 1.439 & 1.441 \\
\hline k &  ---  &  ---  &  ---  &  ---  &  ---  &  ---  & 1.408 \\
\hline
\end{tabular}
\caption{Acenes. HSSH equilibrium C-C bond lengths (in {\AA}).}
\label{tab:Re_ace}
\end{table}

  We compared our HSSH equilibrium C-C bond lengths in acenes to those calculated by Firouzi and Zahedi~\cite{Firouzi:2008} by means of the \emph{ab initio} B3LYP/6-31G(d) approach. As in the case of the phenacenes, see Sec.~\ref{sec:phen}, the agreement is satisfactory, with the exception of the ,,bridge bonds'' (of the symbols d, g, and j), for which the HSSH values are smaller by ca. $0.02$ {\AA}.

  The molecular values of HOMA and its components for acenes are presented in Tabl.~\ref{tab:homa_ace}. As seen, for long acenes HOMA quickly deteriorates, which is consistent with a high chemical reactivity of these molecules.

\begin{table}[!h]
\centering
\begin{tabular}{|c||c|c|c|c|}
\hline Molecule & $\mathrm{R_{\rm av}}$ ({\AA}) & EN & GEO & HOMA\\
\hline
\hline \textbf{Benzene} & 1.397  & 0.021 & 0.000 & 0.979 \\
\hline \textbf{Naphth.} & 1.406  & 0.081 & 0.125 & 0.794 \\
\hline \textbf{30}      & 1.409  & 0.118 & 0.168 & 0.714 \\
\hline \textbf{31}      & 1.411  & 0.141 & 0.181 & 0.678 \\
\hline \textbf{32}      & 1.413  & 0.156 & 0.183 & 0.661 \\
\hline \textbf{33}      & 1.413  & 0.166 & 0.179 & 0.655 \\
\hline \textbf{34}      & 1.414  & 0.174 & 0.173 & 0.653 \\
\hline 
\end{tabular}
\caption{Acenes. Molecular values of $\mathrm{R_{\rm av}}$, EN, GEO, and HOMA.}
 \label{tab:homa_ace}
\end{table}

\newpage

\section{Concluding remarks} \label{sec:concl}

  In a series of papers Tyutyulkov, M\"{u}llen, and their collaborators~\cite{Tyut1998, Dietz2000, Tyut2003, Stay2003} posed a question: ,,is graphene an ultimate large hydrocarbon''? In order to find the answer, they analyzed various aspects: structural (topology), energetic (the energy spectra), and the influence of defects (including these with unpaired spins) and different edge structures. Their findings suggest~\cite{Tyut1998} that the electronic-correlation effects should lead to a nonzero gap in the energy spectrum of PAHs, even in the limit of an infinite molecule (thus, strictly speaking, the metallic limit is not reached). However, the experimental energy gap in the clarene family (\textbf{1} - \textbf{14}), as presented in Fig.~7 of Ref.~\cite{Riegel2010}, indicates that the band gap diminishes quickly with the size of clarene.
  
  Our HSSH study of the equilibrium C-C bond lengths for the PAHs considered in the review paper by Riegel and M\"{u}llen~\cite{Riegel2010} is too limited to attempt an answer to the question of Refs.~\cite{Tyut1998, Dietz2000, Tyut2003, Stay2003}. However,  our elucidation of the modular architecture of some classes of PAHs (clarenes, rylenes, certain subclasses of the K-region PAHs) provides a suggestive indication that the properties of big PAHs may be understood on the basis of the calculations for smaller systems.


\begin{thebibliography}{99}

\bibitem{Stol2020a} L. Z. Stolarczyk, T. M. Krygowski, \textit{J. Phys. Org. Chem.} \textbf{2020}, submitted (Paper I).

\bibitem{Stol2020b} L. Z. Stolarczyk, T. M. Krygowski, \textit{J. Phys. Org. Chem.} \textbf{2020}, submitted (Paper II).

\bibitem{Riegel2010} R. Riegel, K. M\"{u}llen, 
%\emph{Forever young: polycyclic aromatic hydrocarbons as model cases for structural and optical studies} 
\textit{J. Phys. Org. Chem.} \textbf{2010}, \textsl{23}, 315.%-325.

\bibitem{Yama1994} Y. Yamaguchi, Y. Osamura, J. D. Goddard, H. E. Schaefer III,
\textit{A New Dimension to Quantum Chemistry. Analytic Derivative Methods in Ab Initio Molecular Electronic Structure Theory},
Oxford University Press, Oxford, \textbf{1994}.

\bibitem{Couls1940} C. A. Coulson, G. S. Rushbrooke,
%\emph{ Note on the method of molecular orbitals}
\textit{Proc. Cambridge Philos. Soc.} \textbf{1940}, \textsl{36}, 193.%-200.

\bibitem{Krusz1972} J. Kruszewski, T. M. Krygowski, 
%\emph{Definition of Aromaticity Basing on the Harmonic Oscillator Model},
\textit{Tetrahedron Lett.} \textbf{1972}, 3839.%-3842.

\bibitem{Kryg1993} T. M. Krygowski,
%\emph{Crystallographic Studies of Inter- and Intramolecular Interations Reflected in Aromatic Character of $\pi$-Electron Systems},
\textit{J. Chem. Inf. Comput. Sci.} \textbf{1993}, \textsl{33}, 70.%-78.

\bibitem{Kryg1996} T. M. Krygowski, M. Cyra\'{n}ski,
%\emph{Separation of the Energetic and Geometric Contributions to the Aromaticity of $\pi$-Electron carbocyclics},
\textit{Tetrahedron} \textbf{1996}, \textsl{52}, 1713.%-1722.

\bibitem{Cyra1999} M. K. Cyra\'{n}ski, T. M. Krygowski, 
%\emph{Two Sources of the Decrease of Aromaticity:  Bond Length Alternation and Bond Elongation. Part I. An Analysis Based on Benzene Ring Deformations},
\textit{Tetrahedron} \textbf{1999}, \textsl{55}, 6205.%-6210.

%\bibitem{Kryg2014} T. M. Krygowski, H. Szatylowicz, O. A. Stasyuk, J. Dominikowska, M. Palusiak, 
%\emph{Aromaticity from the Viewpoint of Molecular Geometry: Application to Planar Systems}, 
%\textit{Chem. Rev.} \textbf{2014}, \textsl{114}, 63833.%-6422.

\bibitem{Trinajstic:1992} N. Trinajsti{\'c}, \textit{Chemical Graph Theory}, 2nd Ed., CRC Press, Boca Raton (FL), \textbf{1992}.

\bibitem{Clar1972} E. Clar, \textit{The Aromatic Sextet}, Wiley, New York, \textbf{1972}.

\bibitem{Sola2013} M. Sol\`{a}, 
%\emph{Forty years of Clar's aromatic $\pi$-sextet rule}
\textit{Front. Chem.} \textbf{2013}, \textsl{1:22}. %1-8.

\bibitem{scott2015} L. T. Scott,
%\emph{Chemistry at the interior atoms of polycyclic aromatic hydrocarbons}
\textit{Chem. Soc. Rev.} \textbf{2015}, \textsl{44}, 6464.%-6471.

\bibitem{Iyer1997} V. S. Iyer, M. Wehmeier, J. D. Brand, M. A. Keegstra, K. M\"{u}llen,
%\emph{From Hexa-peri-hexabenzocoronene to ``Superacenes''}
\textit{Angew. Chem. Int. Ed. Engl.} \textbf{1997}, \textsl{36}, 1604.%-1607.

\bibitem{Firouzi:2008} R. Firouzi,M. Zahedi,
%\emph{Polyacenes electronic properties and their dependence on molecular size},
\textit{J. Mol. Struct. (Theochem)} \textbf{2008}, \textsl{862}, 7.%-15.

\bibitem{Tyut1998} N. Tyutyulkov, G. Madjarova, F. Dietz, K. M\"{u}llen,  
%\emph{Is 2-D Graphite an Ultimate Large Hydrocarbon? 1. Energy Spectra of Giant Polycyclic Aromatic Hydrocarbons}
\textit{J. Phys. Chem. B} \textbf{1998}, \textsl{102}, 10183.%-10189.

\bibitem{Dietz2000} F. Dietz, N. Tyutyulkov, G. Madjarova, K. M\"{u}llen,
%\emph{Is 2-D Graphite an Ultimate Large Hydrocarbon? II. Structure and Energy Spectra of Polycyclic Aromatic Hydrocarbons with Defects}
\textit{J. Phys. Chem. B} \textbf{2000}, \textsl{104}, 1746.%-1761.

\bibitem{Tyut2003} N. Tyutyulkov, K. M\"{u}llen, M. Baumgarten, A. Ivanova, A. Tadjer,
%\emph{Is 2-D graphite an ultimate large hydrocarbon? III. Structure and energy spectra of large polybenzenoid hydrocarbons with different edge structures}
\textit{Synth. Metals} \textbf{2003}, \textsl{139}, 99.%-107.

\bibitem{Stay2003} A. Staykov, L. Gehrgel, F. Dietz, N. Tyutyulkov, 
%\emph{Is 2-D Graphite an Ultimate Large Hydrocarbon? IV. Structure and Energy Spectra of Polycyclic Aromatic Hydrocarbons with Different Symmetry}
\textit{Z. Naturforsch} \textbf{2003}, \textsl{58b}, 965.%-970.

\end{thebibliography}
\end{document}